\pdfoutput=1
\documentclass[sigplan,screen,11pt,nonacm]{acmart}
\AtBeginDocument{%
  \providecommand\BibTeX{{%
    \normalfont B\kern-0.5em{\scshape i\kern-0.25em b}\kern-0.8em\TeX}}}

\usepackage[capitalize]{cleveref}
\usepackage[frozencache]{minted}
\usepackage{multicol}
\setminted{fontsize=\small}
\newcommand{\haskell}[1]{\mintinline{haskell}{#1}}

\newtheorem{remark}{Remark}[section]

\makeatletter
\AtBeginEnvironment{minted}{\dontdofcolorbox}
\def\dontdofcolorbox{\renewcommand\fcolorbox[4][]{##4}}
\makeatother
\begin{document}

\title{Free Foil: Generating Efficient and Scope-Safe Abstract Syntax}

\author{Nikolai Kudasov}
\email{n.kudasov@innopolis.ru}
\orcid{0000-0001-6572-7292}
\affiliation{%
  \institution{Innopolis University}
  \streetaddress{Universitetskaya, 1}
  \city{Innopolis}
  \state{Tatarstan Republic}
  \country{Russia}
  \postcode{420500}
}
\author{Renata Shakirova}
\email{r.shakirova@innopolis.university}
\affiliation{%
  \institution{Innopolis University}
  \streetaddress{Universitetskaya, 1}
  \city{Innopolis}
  \state{Tatarstan Republic}
  \country{Russia}
  \postcode{420500}
}
\author{Egor Shalagin}
\email{e.shalagin@innopolis.university}
\affiliation{%
  \institution{Innopolis University}
  \streetaddress{Universitetskaya, 1}
  \city{Innopolis}
  \state{Tatarstan Republic}
  \country{Russia}
  \postcode{420500}
}
\author{Karina Tyulebaeva}
\email{k.tyulebaeva@innopolis.university}
\affiliation{%
  \institution{Innopolis University}
  \streetaddress{Universitetskaya, 1}
  \city{Innopolis}
  \state{Tatarstan Republic}
  \country{Russia}
  \postcode{420500}
}

\renewcommand{\shortauthors}{Kudasov, Shakirova, Shalagin, and Tyulebaeva}

\begin{abstract}
% \begin{abstract}
Handling bound identifiers correctly and efficiently is critical
in implementations of compilers, proof assistants, and theorem provers.
When choosing a representation for abstract syntax with binders,
implementors face a trade-off between type safety with intrinsic scoping,
efficiency, and generality.

The ``foil'' by Maclaurin, Radul, and Paszke combines
an efficient implementation of the Barendregt convention with intrinsic scoping
through advanced type system features in Haskell, such as rank-2 polymorphism
and generalized algebraic data types. Free scoped monads of Kudasov, on the other hand,
combine intrinsic scoping with de Bruijn indices as nested data types
with Sweirstra's data types \`a la carte approach
to allow generic implementation of algorithms
such as higher-order unification.

In this paper, we suggest two approaches of making the foil
more affordable. First, we marry free scoped monads with the foil,
allowing to generate efficient, type-safe, and generic abstract syntax representation
with binders for any language given its second-order signature.
Second, we provide Template Haskell functions that allow generating
the scope-safe representation from a na\"ive one.
The latter approach enables us to use existing tools like BNF Converter
to very quickly prototype complete implementation of languages,
including parsing, pretty-printing, and efficient intrinsically scoped abstract syntax.
We demonstrate both approaches using $\lambda\Pi$ with pairs and patterns
as our example object language.

Finally, we provide benchmarks comparing our implementation against the foil,
free scoped monads with nested de Bruijn indices, and some traditional implementations.

  % However, to use the Foil, one still has to write down
  % the (modified) data types for the abstract syntax with binders and manually implement
  % substitution and other functions. Following Swierstra's data types \`a la carte
  % and Kudasov's free scoped monads, we introduce a way to generate the type-safe
  % abstract syntax with binders from a language signature and provide substitution "for free".
  % %Although we leave higher-order unification implementation out of the scope of this paper,
  % %it should also be available for free following the implementation for Kudasov's free scoped monads.
  % Additionally, we provide an extension to the BNF Converter tool allowing to generate
  % parser, pretty-printer, and type-safe abstract syntax for Haskell and Scala backends.
  % We demonstrate the ease of using the Foil with free scoped monads by implementing an
  % interpreter and a typechecker for $\lambda\Pi$, a simple dependently-typed language.
  % Finally, we show that the generated implementation does not have performance penalties,
  % by comparing it with manually implemented Foil, free scoped monads with nested de Bruijn indices,
  % and a na\"ive substitution.

% \keywords{abstract syntax \and second-order \and capture-avoiding substitution \and free monads}
% \end{abstract}

\end{abstract}





\maketitle

\section{Introduction}
\label{section:introduction}

Compilers, theorem provers, and many other formal systems frequently
require handling of scopes, bound names, renaming, and substitution.
In general, manipulating terms with bound names requires capture-avoiding substitution,
which prohibits na\"ive replacement of a variable with an expression.
For instance, in the following example in untyped $\lambda$-calculus,
we cannot simply replace $x$ with $y$ in the last step, since we would mix
bound (local) variable $y$ with free (global) variable $y$:
\[
  (\lambda x. \lambda y. x) y =_{\beta} [x \mapsto y](\lambda y. x)
  \not\equiv \lambda y. y
\]

A na\"ive\footnote{In this paper, we refer to implementations that do not have some scope-safety mechanisms as ``na\"ive''.} representation of untyped lambda terms may look like this in Haskell:

\begin{minted}[texcomments]{haskell}
data Term
  = Var VarIdent      -- $x$
  | App Term Term     -- $(t_1\;t_2)$
  | Lam VarIdent Term -- $\lambda x. t$
\end{minted}

Here, \haskell{VarIdent} is the type of identifiers used for variables
and regardless of specific representation (e.g. \haskell{String} or \haskell{Int}),
capture-avoiding substitution is not trivial to implement and even experts
make mistakes~\cite{Kmett2015_SoH} when using a na\"ive representation like this.

Over the decades, many approaches to capture-avoiding substitution have been put forward.
When implementing a language with binders, one faces a choice of representation,
taking into account efficiency, type safety, and ease of implementation.

\subsection{Existing Approaches}

% \subsubsection{Efficiency.}

Looking for a combination of moderate efficiency and simplicity, a popular approach, at least for small languages, is to use de Bruijn indices~\cite{deBruijn1972}.
However, pure de Bruijn indices suffer from unnecessary shifts and traversals of the subexpressions and many more efficient variations
have been suggested including locally nameless~\cite{Chargueraud2012}, co-de-Bruijn indices~\cite{McBride2018}, ``generalized generalized de Bruijn indices''~\cite{Kmett2015_SoH}, and more.

Some modern compilers and proof assistants rely on an efficient algorithm called ``rapier''~\cite{PeytonJonesMarlow2002}: a technique used by the Glasgow Haskell Compiler,
which is a stateless\footnote{Here we mean that it does not rely on a global state with a fresh name supply.} implementation the Barendregt convention~\cite{Barendregt1985}.
However, the use of scopes in the rapier is not checked, implementing it in practice may lead to bugs as noted by the developers of Dex programming language~\cite{MaclaurinRadulPaszke2023}.

Another way to achieve efficiency is to reuse the mechanism for bound variables
from the host language. This approach is taken by Higher-Order Abstract Syntax (HOAS)~\cite{PfenningElliott1988}
and its variations~\cite{Chlipala2008,WashburnWeirich2008}. This approach also helps with intrinsic scoping,
but has been criticized for the lack of soundness~\cite{FioreSzamozvancev2022} and being difficult to work under binders~\cite{Kmett2015_SoH,Cockx2021_blog}.
% In particular, implementing algorithms, like higher-order unification, that require a lot of matching on the structure of the terms,
% would be inconvenient using this representation.

% \subsubsection{Intrinsic scoping.}

Bird and Paterson's de Bruijn indices as a nested data type~\cite{BirdPaterson1999} parametrize the type of expressions
by a type of (free) variables and rely on polymorphic recursion to extend the scope under binders.
An extension and generalization of this approach is implemented in Haskell in Kmett's \texttt{bound} library~\footnote{\url{https://hackage.haskell.org/package/bound}}~\cite{Kmett2015_SoH}.

Kudasov's ``free scoped monads''~\cite{Kudasov2024freescoped} combine Bird and Paterson's approach with Sweirstra's data types \`a la carte~\cite{Swierstra2008}
and Second-Order Abstract Syntax of Fiore~\cite{FioreHur2010} to generate a scope-safe representation.
With free scoped monads, scope-safe syntax for untyped lambda calculus is generated from a signature
parametrized over \haskell{scope} (a type of scoped subterms $x.t$) and \haskell{term} (a type of regular subterms $t$):
\begin{minted}[texcomments]{haskell}
-- | The (untyped) signature for $\lambda$-terms.
data TermSig scope term
  = App term term   -- $(t_1\;t_2)$
  | Lam scope       -- $\lambda x.t$

-- | $\lambda$-terms with free variables of type 'var'.
type Term var = FreeScoped TermSig var
\end{minted}

Importantly, free scoped monads make it possible to generate capture-avoiding substitution
and provide general recursion over structures with binders, which allows reuse of non-trivial algorithms,
such as higher-order unification~\cite{Kudasov2024freescoped,Kudasov2023}.

In contrast to HOAS, Second-Order Abstract Syntax (SOAS) together with its equational logic and term rewriting
seems to be the correct level of abstraction both for doing metatheory~\cite{FioreSzamozvancev2022,Hamana2017}
and providing generic implementations of algorithms such as higher-order unification~\cite{Kudasov2023}.
While focusing on scope-safety and generality, free scoped monads lack efficiency and adapting them to more efficient
scope-safe representations does not seem immediately straightforward.

The ``foil'' of Maclaurin, Radul, and Paszke~\cite{MaclaurinRadulPaszke2023} offer a combination of efficiency and type-safety.
The foil is effectively a scope-safe implementation of the rapier~\cite{PeytonJonesMarlow2002} that relies on rank-2 polymorphism,
generalized algebraic data types, and phantom type parameters to ensure scope-safety. Users of the foil need to write simple
proofs in a form of method implementations in Haskell for their syntax, ensuring safe renaming.
Scope-safe syntax for untyped lambda calculus may look like this with the foil:
\begin{minted}[texcomments]{haskell}
data Term n where
  Var :: Name n -> Term n           -- $x$
  App :: Term n -> Term n -> Term n -- $(t_1 \; t_2)$
  Lam -- $\lambda x. t$
    :: Binder n l -> Term l -> Term n
\end{minted}
Here the phantom type parameters \haskell{n} and \haskell{l} designate scopes
of the corresponding names, binders and terms. In particular, \haskell{Binder n l}\footnote{Our \haskell{Binder} corresponds to \haskell{NameBinder} from the original paper~\cite{MaclaurinRadulPaszke2023}.}
represents a bound variable that extends scope \haskell{n} to a (larger) scope \haskell{l},
which is the scope of the body of the $\lambda$-abstraction.

\subsection{Contribution}

\begin{figure*}
  \begin{minted}[texcomments]{haskell}
data Term (n :: S) where
  Var      :: Name n -> Term n                          -- $x$
  Pair     :: Term n -> Term n -> Term n                -- $\langle t_1, t_2 \rangle$
  First    :: Term n -> Term n                          -- $\pi_1(t)$
  Second   :: Term n -> Term n                          -- $\pi_2(t)$
  App      :: Term n -> Term n -> Term n                -- $t_1 \; t_2$
  Lam      :: Pattern n l -> Term l -> Term n           -- $\lambda p. t$
  Pi       :: Pattern n l -> Term n -> Term l -> Term n -- $\prod_{p : t_1} t_2$
  Universe :: Term n                                    -- $\mathcal{U}$
  \end{minted}
  \caption{The foil representation for terms in $\lambda\Pi$ with pairs and patterns.
  \haskell{Term} and \haskell{Name} are parametrized by a phantom type parameter that indicates the scope (untyped context) of the term.
  \haskell{Pattern} is parametrized by two phantom parameters: first parameter indicates the smaller outer scope
  while the second parameter indicates the larger inner scope, extended with the variables bound in the pattern.
  }
  \label{figure:foil-lambda-pi}
\end{figure*}

In this paper, we introduce two approaches to the generation of
a scope-safe abstract syntax, based on the foil~\cite{MaclaurinRadulPaszke2023}.
Our first approach relies on a variation of free scoped monads~\cite{Kudasov2024freescoped},
providing general recursion schemes and support for data types \`a la carte~\cite{Swierstra2008}.
Our second approach instead relies on metaprogramming (via Template Haskell~\cite{SheardPeytonJones2002})
to convert a na\"ive representation into a scope-safe one.

The first approach does not rely on any metaprogramming and works well when
implementing generic recursion schemes over abstract syntax with binders.
The second approach couples well with parser generators, like BNF Converter~\cite{BNFConverter2004},
and supports non-trivial patterns for binders.

The outline of the rest of this paper is as follows:

\begin{enumerate}
  \item In \cref{section:foil-review}, we briefly review the foil~\cite{MaclaurinRadulPaszke2023}
  and extend it with support for non-trivial patterns.
  \item In \cref{section:free-foil}, we introduce a variant of free scoped monads, which we call ``free foil'',
  that relies on the foil (replacing Bird and Patterson's de Bruijn notation as nested data types).
  \item In \cref{section:template-haskell}, we describe our metaprogramming approach to
  the generation of the abstract syntax with the foil (not free foil), from a na\"ive representation.
  We require clear separation of syntactic categories (terms, scopes, patterns, variables),
  which simplifies the analysis of user definitions and, we believe, is reasonable to ask of the users.
  % \item In \cref{section:example}, we demonstrate how free foil can be used
  % to implement $\lambda\Pi$, a small dependently typed programming language.
  \item In \cref{section:benchmarks}, we extend the Weirich's bechmarks~\cite{lambda-n-ways}
  for implementations of untyped lambda calculus and compare the performance of free foil
  against some traditional implementations as well as against the foil~\cite{MaclaurinRadulPaszke2023}
  and free scoped monads~\cite{Kudasov2024freescoped}.
  % \item In \cref{section:haskell-vs-scala}, we discuss some of the technical differences between
  % implementations of the Free Foil in Haskell and Scala.
\end{enumerate}

The code with all definitions presented in this paper with examples
is available at \href{https://github.com/fizruk/free-foil}{github.com/fizruk/free-foil}.
The benchmarks presented in \cref{section:benchmarks} are available at \href{https://github.com/KarinaTyulebaeva/lambda-n-ways}{github.com/KarinaTyulebaeva/lambda-n-ways}.

% Examples of todos:

% \nikolai{here be dragons}
% \renata{here be dragons}
% \egor{here be dragons}
% \karina{here be dragons}

\section{The Foil}
\label{section:foil-review}

In this section, we review the major components of the foil~\cite{MaclaurinRadulPaszke2023},
as we reuse most of their types and type class machinery in the following sections.
We also introduce an extra type class to assist with the support of non-trivial patterns,
which can be seen as a bundle of zero or more binders.
We demonstrate the foil's approach by using the foil's representation for dependently typed lambda calculus with (non-dependent) pairs.

\subsection{Reviewing the Foil}

\begin{figure*}
\begin{minted}{haskell}
instance Sinkable Term where
  sinkabilityProof :: (Name n -> Name l) -> Term n -> Term l
  sinkabilityProof rename = \case
    Var x -> Var (rename x)
    Pair l r -> Pair (sinkabilityProof rename l) (sinkabilityProof rename r)
    First x -> First (sinkabilityProof x)
    Second x -> Second (sinkabilityProof x)
    App fun arg -> App (sinkabilityProof rename fun) (sinkabilityProof rename arg)
    Lam pat body -> coSinkabilityProof rename pat $ \rename' pat' ->
      Lam pat' (sinkabilityProof rename' body)
    Pi pat argType retType -> coSinkabilityProof rename pat $ \rename' pat' ->
      Pi pat' (sinkabilityProof rename argType) (sinkabilityProof rename' retType)
    Universe -> Universe
\end{minted}
\caption{Sinkability proof for $\lambda\Pi$ terms.
This definition ensures that renaming is scope-safe for \haskell{Term}
and only serves as a formal justification to use a zero cost function \haskell{sink}
instead of \haskell{sinkabilityProof} in the user code.}
\label{figure:foil-sinkability-lambda-pi}
\end{figure*}

The main idea behind the foil is to annotate every expression with
a phantom type parameter to keep track of the (abstract representation of the) in-scope variables that \emph{may}
be present within the expression. This parameter makes sure that we
perform the necessary scope extensions and rename bound variables, while enabling the runtime to
sometimes skip the checks or renaming entirely, when it is known to be fine.
The foil can be seen as an application of the technique called ``the ghosts of departed proofs''~\cite{Noonan2018}.

\cref{figure:foil-lambda-pi} demonstrates the foil representation for $\lambda\Pi$ with pairs.
The \haskell{n} type variable is a phantom parameter~\footnote{A \emph{phantom} type parameter is one that does not affect the representation of the parametrized type.}
that corresponds with the scope to which the term belongs. Note that this parameter is abstract,
and usually we cannot extract the actual set of variable names given a particular type \haskell{n}.
Instead, scope safety is ensured by relying on the limited information about relationships between different scopes.
For example, constructors \haskell{Lam} and \haskell{Pi} both introduce local variables bound by a pattern
that extends scope \haskell{n} to scope \haskell{l}. Then, the body of a lambda abstraction is a term in the extended scope \haskell{l},
while the overall lambda abstraction is in scope \haskell{n}. In this example, we use \haskell{Pattern} which is a generalization of
single variable binder used in the original paper and is discussed below.

In the following list we present the core types and type classes which constitute the foil.
Note that every type is scope-indexed, and the underlying constructors often start with \haskell{Unsafe} prefix.
However, these constructors are not exposed to the users of the foil.

\begin{enumerate}
  \item \haskell{S} (for ``Scope'') is a kind\footnote{A \emph{kind} is a type of a type or a type constructor.}
  of scope-indexing types. The only concrete scope that is available to the users is the empty scope \haskell{VoidS :: S}.
  All other scopes are abstract and represented in the code as type variables like \haskell{n :: S}.

  \item \haskell{Name n} is the type of identifiers in the scope \haskell{n :: S}.
  In the original paper as well as in this paper,
  machine-sized integer is used for the underlying representation.

  \item \haskell{Scope n} is the type of scopes (``sets'' of variables) corresponding to \haskell{n}.
  In the foil, following the rapier~\cite{PeytonJonesMarlow2002}, a purely functional
  integer map~\cite{OkasakiGill1998} is used for fast lookups and extensions of the scope.
  We expect \haskell{Scope n} to be a singleton and provide exactly the names in \haskell{Name n}:
  \begin{enumerate}
      \item if \haskell{s1 :: Scope n} and \haskell{s2 :: Scope n}, then \haskell{s1 == s2};
      \item if \haskell{s :: Scope n} and \haskell{x : Name n}, then \haskell{x `member` s}.
  \end{enumerate}

  \item \haskell{Binder n l} is the type of a single identifier that extends
  scope \haskell{n} to scope \haskell{l}. Binders are also names:
  \begin{minted}{haskell}
nameOf :: Binder n l -> Name l
  \end{minted}

  Binders can extend scopes:
  \begin{minted}{haskell}
extendScope
  :: Binder n l -> Scope n -> Scope l
  \end{minted}

  To prevent leaking variables from extended scopes, the foil
  relies on rank-2 polymorphism:
  \begin{minted}{haskell}
withFreshBinder
  :: Scope n
  -> (forall l. Binder n l -> r)
  -> r
  \end{minted}
  Here, the extended scope \haskell{l} is bound by a \haskell{forall},
  making sure that the caller of this function cannot know
  the extended scope in advance.

  \item \haskell{class Ext n l} is a type class without any methods that
  ensures that scope \haskell{l} extends scope \haskell{n}.

  \item \haskell{class Distinct n} is a type class without any methods
  that ensures that scope \haskell{n} contains distinct variables.
  Both \haskell{Ext n l} and \haskell{Distinct n l} are never supposed to be
  instantiated by the user, but are implied in rank-2 polymorphic functions.

  \item \haskell{class Sinkable (e :: S -> *)} is a type class with
  only one method renames variables in an expression given a renaming function,
  possibly changing scopes:
  \begin{minted}{haskell}
sinkabilityProof
  :: (Name n -> Name l) -> e n -> e l
  \end{minted}
  An example of such a proof for $\lambda\Pi$ is given in~\cref{figure:foil-sinkability-lambda-pi}.
  It is crucial that \haskell{sinkabilityProof} merely provides
  a proof that such renaming is safe. Given the specific representation
  of names in the foil, renaming turns out to always be the identity
  on the underlying representation, warranting the following efficient
  alternative to be used once the proof is provided by the user:
  \begin{minted}{haskell}
sink
  :: (Sinkable expr, Distinct n, Ext n l)
  => expr n -> expr l
sink = unsafeCoerce
  \end{minted}
\end{enumerate}

\begin{figure*}
\begin{minted}{haskell}
substTerm :: Distinct o => Scope o -> Subst Term i o -> Term i -> Term o
substTerm scope subst = \case
  Var x    -> lookupSubst subst x
  Pair l r -> Pair (substTerm scope subst l) (substTerm scope subst r)
  First x  -> First (substTerm scope subst x)
  Second x -> Second (substTerm scope subst x)
  App f x  -> App  (substTerm scope subst f) (substTerm scope subst x)
  Lam pat body -> withPattern scope pat subst $ \pattern' subst' scope' ->
    let body' = substTerm scope' subst' body
    in Lam pattern' body'
  Pi pat typ body -> withPattern scope pat subst $ \pattern' subst' scope' ->
    let body' = substTerm scope' subst' body
        typ' = substTerm scope subst typ
    in Pi pattern' typ'  body'
  Universe -> Universe
\end{minted}
\caption{Scope-safe implementation of the substitution for the foil representation of terms in $\lambda\Pi$ with pairs and patterns.
The foil helps maintain the correct implementation, making sure binders and scopes are handled properly.}
\label{figure:foil-patterns-substitution}
\end{figure*}

The function \haskell{sink} demonstrates one of the main advantages of using a
name-based representation for abstract syntax with binders, as opposed to de Bruijn indices.
Indeed, for de Bruijn indices, in place of \haskell{sink}, we would have a function
that would increment all indices in a subexpression, to lift it to an extended scope.
Although some approaches can optimize their perfomance of such an increment, it is never free.
On the other hand, \haskell{sink} has no runtime cost and, since the type parameters \haskell{n} and \haskell{l}
are phantom and the foil controls the underlying representation of names,
the unsafe coercion here is actually completely safe.

\subsection{Safe Substitutions}

\begin{figure*}
\begin{minted}{haskell}
class CoSinkable (pattern :: S -> S -> *) where
  coSinkabilityProof
    :: (Name n -> Name n')
    -> pattern n l
    -> (forall l'. (Name l -> Name l') -> pattern n' l' -> r) -> r

instance CoSinkable Binder where
  coSinkabilityProof _rename (UnsafeBinder name) cont =
    cont unsafeCoerce (UnsafeBinder name)

extendRenaming
  :: CoSinkable pattern
  => (Name n -> Name n')
  -> pattern n l
  -> (forall l'. (Name l -> Name l') -> pattern n' l' -> r) -> r
extendRenaming _rename pattern cont = cont unsafeCoerce (unsafeCoerce pattern)
\end{minted}
\caption{The \haskell{CoSinkable} type class, its instance for \haskell{Binder},
and a general version of \haskell{extendRenaming}.}
\label{figure:cosinkable}
\end{figure*}

A substitution is a mapping from names in a given scope to expressions.
The underlying representation in the foil relies on integer maps
for efficiency:
\begin{minted}{haskell}
data Subst (e :: S -> *) (i :: S) (o :: S)
  = UnsafeSubst
    (forall n. Name n -> e n) (IntMap (e o))
\end{minted}

Here, \haskell{Subst} is parametrized by the type constructor for expressions \haskell{e},
the scope \haskell{i} prior to the substitution, and scope \haskell{o}
that we obtain after performing the substitution. For example, \haskell{Subst Term n VoidS}
is a type of substitutions for $\lambda\Pi$ (as in~\cref{figure:foil-lambda-pi}) that
defines how names in scope \haskell{n} are mapped to closed\footnote{A \emph{closed} term is a term without free variables.}
$\lambda\Pi$ terms.

The constructor \haskell{UnsafeSubst} is not available to the users of the foil,
instead the substitutions are constructed using the following combinators:

\begin{minted}{haskell}
idSubst
  :: (forall n. Name n -> e n) -> Subst e i i

addSubst
  :: Subst e i o    -- substitution for scope i
  -> Binder i i'    -- new name in scope i'
  -> e o            -- mapping for the new name
  -> Subst e i' o   -- substitution for scope i'

addRename
  :: Subst e i o    -- substitution for scope i
  -> Binder i i'    -- new name in scope i'
  -> Name o         -- mapping for the new name
  -> Subst e i' o   -- substitution for scope i'
\end{minted}

The first component of the substitution is an injection of variables into expressions,
which is required to retain unmapped variables.
For example, in case of $\lambda\Pi$ representation in~\cref{figure:foil-lambda-pi}
this component is supposed to be exactly \haskell{Var}.
While it is possible to move this component into a typeclass,
here we prefer to follow the original description of the foil.

That said, in \cref{section:free-foil}, we introduce a type that allows us to have one \haskell{Var}
constructor for different languages, improving \haskell{Subst} to a mere \haskell{newtype}-wrapper around \haskell{IntMap (e o)}
without using typeclasses.
We note also that the authors of the foil implement a similar and more advanced optimization
in their implementation of Dex\footnote{In some places, they use \haskell{SubstFrag} type which is essentially a newtype for the \haskell{IntMap}.
In particular, it can be used directly for \haskell{Subst}, see \url{https://github.com/google-research/dex-lang/blob/438d9574745d7a5b724fb2aa41effa6255c74eda/src/lib/Name.hs\#L52-L59}}:
since they implement the foil for a specific language,
they drop this dependency, and they also introduce a few coercions between scopes and substitutions.

Substitutions are also \haskell{Sinkable}:
\begin{minted}{haskell}
instance Sinkable e => Sinkable (Subst e i) where
  sinkabilityProof rename (UnsafeSubst f env) =
    UnsafeSubst f (fmap (sinkabilityProof rename) env)
\end{minted}

Given a substitution \haskell{subst :: Subst e i o},
we can sink it to \haskell{Subst e i o'}
for any scope \haskell{o'} that extends \haskell{o}.
This together with \haskell{addSubst} and \haskell{addRename} enables the use
of substitutions under binders.

% \begin{minted}{haskell}
%   substituteTerm
%     :: Distinct o => Scope o -> Subst Term i o -> Term i -> Term o
%   substituteTerm scope subst = \case
%     Var x -> lookupSubst subst x
%     App fun arg ->
%       App (substituteTerm scope subst fun)
%           (substituteTerm scope subst arg)
%     Lam pat body -> withRefreshed
%   \end{minted}

%   \subsubsection{Subst Invariant} \

%   Here we are tackling the final invariant, ensuring accurate substitution and error prevention, for which the creation of a \haskell{Subst} type class is necessary.

%   The \haskell{Subst} requires mapping input names to the terms being substituted, with a performance optimization for scenarios where a name is being changed to itself, commonly occurring within a local scope and under a non-shadowing binder at runtime.

%    If a \haskell{Subst e i o} exists, then it gives semantics to every \haskell{name :: Name i}, and \haskell{lookupSubst} implements those semantics.

%   \textit{Distinctness Invariant:}
%   \begin{enumerate}
%     \item If name was last added with an \haskell{addSubst} whose third argument was \haskell{e}, then \haskell{lookupSubst} returns \haskell{sink e}.
%     \item If name was last added with an \haskell{addRename} whose third argument was \haskell{name'}, then \haskell{lookupSubst} returns \haskell{f name'}.
%     \item Otherwise, \haskell{name} was present in the root scope at which we called \haskell{idSubst} and \haskell{lookupSubst} returns \haskell{f (unsafeCoerce name)}.
%   \end{enumerate}

% \subsection{Safe Capture-Avoiding Substitution}

In~\cref{figure:foil-patterns-substitution} we present an implementation of capture-avoiding substitution for $\lambda\Pi$ terms defined in~\cref{figure:foil-lambda-pi}.
Here, \haskell{withPattern} is a generalization of \haskell{withRefreshed}~\cite[Figure~6]{MaclaurinRadulPaszke2023}
for patterns and is presented in the next section where we extend the foil with patterns.

Note that the foil makes sure our definition is correct.
For example, if we do not apply \haskell{substTerm} for one of the arguments in,
for instance, \haskell{App}, the function will not type check.
Similarly, without refreshing names using \haskell{withPattern} for \haskell{Lam} or \haskell{Pi},
it is impossible to go under a binder, since the body's type is different.

\subsection{Extending the Foil with Patterns}

\begin{figure*}
\begin{minted}{haskell}
data Pattern (n :: S) (l :: S) where
  PatternWildcard :: Pattern n n
  PatternVar      :: Binder n l -> Pattern n l
  PatternPair     :: Pattern n k -> Pattern k l -> Pattern n l

instance CoSinkable Pattern where
  coSinkabilityProof rename pattern cont = case pattern of
    PatternWildcard -> cont rename PatternWildcard
    PatternVar binder ->
      coSinkabilityProof rename binder $ \ rename' binder' ->
        cont rename' (PatternVar binder')
    PatternPair l r ->
      coSinkabilityProof rename l $ \ renameL l' ->
        coSinkabilityProof renameL r $ \ renameR r' ->
          cont renameR (PatternPair l' r')

withPattern
:: DExt s o
=> Scope -> Pattern n i -> Subst Term n s
-> (forall o'. Distinct o' => Pattern o o' -> Subst Term i o' -> Scope o' -> r)
-> r
withPattern scope pat subst cont = case pat of
  PatternWildcard -> cont PatternWildcard (sink subst) scope
  PatternVar var -> withRefreshed scope (nameOf var) $ \ var' ->
    let subst' = addRename (sink subst) var (sink (nameOf var'))
        scope' = extendScope var' scope
    in cont (PatternVar var') subst' scope'
  PatternPair p1 p2 ->
    withPattern scope p1 subst $ \p1' subst' scope' ->
      withPattern scope' p2 subst' $ \p2' subst'' scope'' ->
        cont (PatternPair p1' p2') subst'' scope''
\end{minted}
\caption{The foil representation for a complex pattern type with corresponding \haskell{CoSinkable} instance and \haskell{withPattern}.}
\label{figure:pattern}
\end{figure*}

In the original paper, \haskell{Binder n l} extends the scope \haskell{n} with exactly one local variable,
arriving at the scope \haskell{l}.
However, in most languages, patterns are allowed to bind zero or more variables.
To represent patterns more faithfully, we add one more typeclass to the foil,
called \haskell{CoSinkable} and presented in~\cref{figure:cosinkable}.

\haskell{CoSinkable}, as the name suggests, is dual to \haskell{Sinkable}
in the following sense. While \haskell{sinkabilityProof} applies a given renaming
to the free variables in an expression, \haskell{coSinkabilityProof} makes
sure that variables in patterns are refreshed properly to avoid name capture.
The type signature of \haskell{coSinkabilityProof} is more complicated as it
is given in a continuation-passing style and relies on rank-2 polymorphism
to abstract away the new extended scope \haskell{l'}.
However, it can also be understood as follows: given a renaming function
and a pattern, \haskell{coSinkabilityProof} provides a renaming function
for the variables bound by the pattern, and a refreshed version of the pattern.

The function \haskell{coSinkabilityProof} generalizes a function that is originally defined
only on \haskell{Binder}~\cite[Appendix A]{MaclaurinRadulPaszke2023}.
Naturally, we use the same implementation to define the instance of \haskell{CoSinkable Binder},
as we also do not know of an implementation for \haskell{Binder} that does not rely on \haskell{unsafeCoerce}.
That said, once we have the instance for \haskell{Binder}, instances for more complex patterns
can be built safely, as we will see later in this section.

Similarly to the \haskell{sinkabilityProof}, we only require \haskell{coSinkabilityProof}
for typechecking, and we can always use an efficient version \haskell{extendRenaming},
presented in~\cref{figure:cosinkable}, in the implementation of the object language.

% \nikolai{We should try to find a safe implementation of this instance.}

To complete our $\lambda\Pi$ example we define a type with patterns for wildcards, variables,
and pairs and prove that it is \haskell{CoSinkable}. The definitions are presented in~\cref{figure:pattern}.
Note that \haskell{PatternWildcard} has the same ``outer'' and ``inner'' scopes,
while in the \haskell{PatternPair} the type parameter \haskell{k} is
existentially quantified as it does not appear in the resulting type.

% \nikolai{Do we require extra laws? E.g. \haskell{coSinkabilityProof id pattern k = k id pattern}?}

For substitution presented in~\cref{figure:foil-patterns-substitution}, we introduce a generalization of \haskell{withRefreshed}~\cite[Figure~6]{MaclaurinRadulPaszke2023} for patterns.
In~\cref{figure:foil-patterns-substitution}, \haskell{withPattern} traverses over pattern, extending given \haskell{subst} and \haskell{scope} with binders defined within the pattern.
% \haskell{withRefreshed} cannot be formulated in terms of \haskell{coSinkabilityProof}
% and has to be implemented for each pattern type, even though it has a similar structure.

% Remember, that \haskell{sinkabilityProof} here is required only for typechecking
% and the actual implementation used for \haskell{sink} is \haskell{unsafeCoerce} with zero runtime cost.

% \subsection{Comparison with Rapier}

% The foil itself is a kind of improvement on Rapier, bringing all its operations to the type level using the strong type system that is present in Haskell and Scala.

% The Foil introduces a number of invariants that we want to use when name freshing, substitution, and scope expansion and expresses these invariants at the type level, thus making substitution operations safer.
% One challenge of utilizing Rapier is its lack of type restrictions, which can lead to the compilation of erroneous programs that may remain undetected for an extended period of time.
% Foil invariants serve as a powerful safeguard, significantly raising the bar for bug introduction. This characteristic stands as the greatest strength of Foil.

% \subsection{Supporting patterns}

% To support

\section{Free Foil}
\label{section:free-foil}

\begin{figure*}
\begin{minted}{haskell}
data ScopedAST sig n where
  ScopedAST :: Binder n l -> AST sig l -> ScopedAST sig n

data AST sig n where
  Var  :: Name n -> AST sig n
  Node :: sig (ScopedAST sig n) (AST sig n) -> AST sig n

instance Bifunctor sig => Sinkable (AST sig) where
  sinkabilityProof :: (Name n -> Name l) -> AST sig n -> AST sig l
  sinkabilityProof rename = \case
    Var name -> Var (rename name)
    Node node -> Node (bimap f (sinkabilityProof rename) node)
    where
      f (ScopedAST binder body) =
        extendRenaming rename binder $ \rename' binder' ->
          ScopedAST binder' (sinkabilityProof rename' body)

substitute :: (Bifunctor sig, Distinct o) => Scope o -> Subst (AST sig) i o -> AST sig i -> AST sig o
substitute scope subst = \case
  Var name -> Foil.lookupSubst subst name
  Node node -> Node (bimap f (substitute scope subst) node)
  where
    f (ScopedAST binder body) =
      Foil.withRefreshed scope (Foil.nameOf binder) $ \binder' ->
        let subst' = Foil.addRename (Foil.sink subst) binder (Foil.nameOf binder')
            scope' = Foil.extendScope binder' scope
            body' = substitute scope' subst' body
        in ScopedAST binder' body'
\end{minted}
\caption{The Free Foil: \haskell{AST sig n} is the intrinsically scoped abstract syntax
representation based on the foil, freely generated from a signature \haskell{sig} of an object language.
Whenever \haskell{sig} is a \haskell{Bifunctor}, the generated syntax is sinkable
and has capture-avoiding substitution defined for it. The \haskell{Bifunctor} here allows to apply transformations to scoped and regular subterms (separately) via \haskell{bimap}.}
\label{figure:free-foil}
\end{figure*}

In this section, we develop a variant of free scoped monads~\cite{Kudasov2024freescoped}
but relying on the foil~\cite{MaclaurinRadulPaszke2023} for intrinsic scoping instead of Bird and Patterson's de Bruijn notation
as nested data types~\cite{BirdPaterson1999}.

Originally, free scoped monads are used to generate Second-Order Abstract Syntax~\cite{FioreHur2010}, which includes parametrized metavariables that play an important role in the equational logic, term rewriting, and unification.
However, in the context of this paper, parametrized metavariables are not important, so we omit them.
Note that parametrized metavariables can be easily restored with the data types \`a la carte approach~\cite[\S 2.3]{Kudasov2024freescoped}.

The main definitions of the free foil are given in~\cref{figure:free-foil}.
The \haskell{ScopedAST} data type generalizes \haskell{LamExpr} from the foil~\cite[\S 3.2]{MaclaurinRadulPaszke2023}
and introduces a single binder to be used in a subterm.
The signature parameter \haskell{sig} is assumed to be a \haskell{Bifunctor}\footnote{A two-parameter type constructor \haskell{f} is a \haskell{Bifunctor} if it supports mapping over any or both of its parameters. This can be seen as an extension of a \haskell{Functor} to type constructors with two parameters.}
with the two type parameters corresponding to scoped subterms and regular subterms respectively.
The \haskell{AST} data type then freely generates the recursive abstract syntax
and adds variables.

The substitution can be defined in the general case, given a \haskell{Sinkable} instance for the generated syntax.
The main differences of the freely generated substitution from the original foil is that now refreshing the scope is done within applying \haskell{bimap} for
a node in the AST, and not when substituting into \haskell{Lam} expression specifically.
Thus, it is now generalized for any term which expands the scope by introducing a local name,
allowing for capture-avoiding substitution on any syntactic construction (such as \haskell{let}-bindings and \haskell{for}-loops).

\begin{remark}
While \haskell{substitute} and \haskell{Var} are reminiscent of the bind and unit of a monad,
\haskell{AST sig} does not fit Haskell's \haskell{Monad} typeclass:
although \haskell{Subst (AST sig) i o} can be replaced with a function \haskell{Name i -> AST sig o},
the problem is that \haskell{AST sig} is parametrized by a phantom scope parameter (of kind \haskell{S}), not a type (of kind \haskell{*}),
and there is also an extra \haskell{Scope o} parameter.
However, it turns out that \haskell{AST' sig} is a \haskell{Name}-relative monad~\cite{AltenkirchChapmanUustalu2015}, where
\haskell{newtype AST' sig n = AST' (Scope n -> AST sig n)}.
\end{remark}

\begin{figure*}
\begin{minted}[texcomments]{haskell}
data LambdaPiF scope term   -- Signature of $\lambda\Pi$ (without pair).
  = AppF term term    -- $t_1 \; t_2$
  | LamF scope        -- $\lambda x. t$
  | PiF term scope    -- $\prod_{x : A} B(x)$
  | UniverseF         -- $\mathcal{U}$

data PairF scope term       -- Signature for pairs and projections.
  = PairF term term   -- $\langle t_1 t_2 \rangle$
  | FirstF term       -- $\pi_1(t)$
  | SecondF term      -- $\pi_2(t)$

-- | Sum of signatures, enabling data type \`a la carte approach.
data (f :+: g) scope term = InL (f scope term) | InR (g scope term)

type Term n = AST (LambdaPiF :+: PairF) n

pattern App fun arg     = Node (InL (AppF fun arg))
pattern Lam binder body = Node (InL (LamF (ScopedAST binder body)))
pattern Pi binder a b   = Node (InL (PiF a (ScopedAST binder b)))
pattern Universe        = Node (InL UniverseF)
pattern Pair l r        = Node (InR (PairF l r))
pattern First t         = Node (InR (FirstF t))
pattern Second t        = Node (InR (SecondF t))

whnf :: Distinct n => Scope n -> Term n -> Term n
whnf scope = \case
  First t -> case whnf t of
    Pair l _r -> whnf l
    t' -> First t'
  Second t -> case whnf t of
    Pair _l r -> whnf r
    t' -> Second t'
  App f x -> case whnf scope f of
    Lam binder body ->
      let subst = addSubst identitySubst binder x
      in whnf scope (substitute scope subst body)
    f' -> App f' x
  t -> t
\end{minted}
\caption{Generated syntax and pattern synonyms for $\lambda\Pi$ using the free foil,
and an implementation of weak head normal form evaluation for it.}
\label{figure:free-foil-lambda-pi}
\end{figure*}

It is also remarkable that the foil's \haskell{Subst} data structure and all invariants and operations associated with it are generic enough to work with our newly defined syntax generating data type out of the box. The only requirement is to define an instance of \haskell{Sinkable} for it.

And even in this instance, \haskell{extendRenaming}, originally designed for sinking \haskell{Lam} terms, is directly used here for a generic node with scoped subterms. Therefore, the initial design of the foil is suitable for adapting it to freely generated abstract syntax, mainly due to
generic safe sinking~\cite[\S 3.5]{MaclaurinRadulPaszke2023}.

With this generic structure of subterms, we can now instantiate such structure for any object language,
for instance, already mentioned $\lambda\Pi$ with pairs, and get type-safe and efficient substitution for it for free.
Similarly to original free scoped monads~\cite[\S 2.2]{Kudasov2024freescoped},
we can introduce pattern synonyms for easy pattern matching for the generated AST.
The signature and pattern synonyms for $\lambda\Pi$, as well as a definition of
weak head normal form evaluation are given in~\cref{figure:free-foil-lambda-pi}.
In this presentation, we split $\lambda\Pi$ into two signatures to demonstrate
the data types \`a la carte approach.

Note the explicit passing of \haskell{scope} in the definition of \haskell{whnf} in~\cref{figure:free-foil-lambda-pi},
which is a sign of using the Barendregt convention, which the rapier and the foil follow.
Other that this minor adjustment, the implementation of \haskell{whnf} is straightforward
and Haskell's type system ensures that we do not forget handling of the scopes.
As an advantage over the original foil implementation, substitution is provided for free,
and is available for any target language provided its signature.

\section{Foil via Template Haskell}
\label{section:template-haskell}

In this section, we describe a metaprogramming approach to the generation of
abstract syntax representation based on the foil~\cite{MaclaurinRadulPaszke2023},
relying on Template Haskell~\cite{SheardPeytonJones2002}.

The motivation for such an approach is twofold. First, it is convenient
to automate transition from a na\"ive representation to the foil, lowering
the entry threshold for developers, less familiar with the extensions such as rank-2 polymorphism
and generalized algebraic data types.
Once the suitable definitions are generated, the developers should only
``follow the types''\footnote{``Follow the types'' is the phrase often used by Haskellers,
referring to the fact that once the types are written down, the implementation conveniently follows from it.
A similar philosophy is embraced by the Type-Driven Development~\cite{Brady2017}.},
without having to figure out the correct use of these features to define the representation in the first place.
Second, existing parser generator tools such as BNF Converter~\cite{BNFConverter2004,BNFCMeta} are able
to generate (na\"ive) abstract syntax representation based on a grammar expressed in an extended BNF.
Generating proper conversion functions from that to the foil allows going straight
from an extended BNF grammar to the scope-safe representation, significantly speeding up
prototype development.

\subsection{Classifying User Syntax}

To be able to generate the foil representation for the object language,
we need to understand where variable identifiers can be found in the na\"ive
representation and in what capacity they are being used (as a variable or as a binder).
Many design choices are available here, especially if explicit annotations are allowed.
However, in this paper, we aim for the least invasive approach,
requiring no special annotations, except asking to organize the user's representation
into four (classes of) types:
\begin{enumerate}
  \item Type of \emph{variable identifiers}. Identifiers can have arbitrary representation,
  but to avoid unnecessary collision with other data in the syntax representation,
  we prefer the user to use a wrapper type (e.g. using \haskell{newtype}), for instance:
\begin{minted}{haskell}
newtype VarIdent = VarIdent String
\end{minted}

  \item Type of \emph{patterns (binders)}. This type corresponds to the possible
  patterns allowed in binding constructions. In the simplest case, this can be a wrapper
  around a variable identifier. But it can also be something more advanced,
  allowing wildcards or complex patterns, e.g.:
\begin{minted}{haskell}
data Pattern
  = PatternWildcard
  | PatternVar VarIdent
  | PatternPair Pattern Pattern
\end{minted}

  \item Type of \emph{scoped terms}. Normally, this should simply be a wrapper
  around a term type (described below):
\begin{minted}{haskell}
newtype ScopedTerm = ScopedTerm Term
\end{minted}

  \item Type of \emph{terms} that relies on the three types above to
  mark the use of patterns and scopes. In this paper, we additionally assume
  that each term constructor has at most one pattern and all scoped subterms
  may have access to the variables introduced by that pattern:
\begin{minted}[texcomments]{haskell}
data Term
  = Var VarIdent                -- $x$
  | Pair Term Term              -- $\langle t_1, t_2 \rangle$
  | First Term                  -- $\pi_1(t)$
  | Second Term                 -- $\pi_1(t)$
  | App Term Term               -- $(t_1 \; t_2)$
  | Lam Pattern ScopedTerm      -- $\lambda x. t$
  | Pi Pattern Term ScopedTerm  -- $\prod_{x : t_1} t_2$
  | Universe                    -- $\mathcal{U}$
\end{minted}
\end{enumerate}

% \nikolai{Do we allow term types used in patterns?
% E.g. for type ascriptions in patterns, or for multiple \haskell{let}-bindings?
% How do we deal with lists or patterns?}

In this paper, we only handle one type per class, however, our Template Haskell
implementation should scale naturally to supporting arbitrary number of types per each of the classes
for extra flexibility.

\subsection{Generating the Foil}

\begin{figure*}
\begin{minted}{haskell}
data FoilTerm n where
  FoilVar   :: Foil.Name n -> FoilTerm n
  FoilPair  :: FoilTerm n -> FoilTerm n -> FoilTerm n
  FoilApp   :: FoilTerm n -> FoilTerm n -> FoilTerm n
  FoilLam   :: FoilPattern n l -> FoilTerm l -> FoilTerm n
  FoilPi    :: FoilPattern n l -> FoilTerm n -> FoilTerm l -> FoilTerm n

data FoilPattern n l where
  FoilPatternWildcard :: FoilPattern n n
  FoilPatternVar :: Foil.Binder n l -> FoilPattern n l
  FoilPatternPair :: FoilPattern n l1 -> FoilPattern l1 l -> FoilPattern n l

data FoilScopedTerm n where
  FoilScopedTerm :: FoilTerm n -> FoilScopedTerm n
\end{minted}
\caption{Foil representation for $\lambda\Pi$ with pairs and patterns, generated via \haskell{mkFoilData}.}
\label{figure:foil-data}
\end{figure*}

If the user-defined syntax complies with our minimal expectations,
we are able to generate the corresponding foil representation together with conversion functions.
More specifically, given the names of the types for variable identifiers, patterns, terms, and scoped terms, we provide the following Template Haskell functions:
\begin{enumerate}
  \item \haskell{mkFoilData} generates the foil equivalent for patterns, terms, and scoped terms\footnote{Variable identifiers in foil have a fixed representation as \haskell{Name n} and \haskell{Binder n l}.};
  \item \haskell{mkToFoil} generates conversion functions for patterns, terms, and scoped terms to their respecitive foil data types;
  \item \haskell{mkFromFoil} generates conversion functions for patterns, terms, and scoped terms from their respecitive foil data types;
  \item \haskell{mkInstancesFoil} generates instances for the \haskell{Sinkable} type class for terms and scoped terms
  as well as instances for the \haskell{CoSinkable} type class for patterns;
\end{enumerate}

The generated scope-safe representation provided by \haskell{mkFoilData} is shown in~\cref{figure:foil-data}.
When generating the foil data types, we follow a simple conversion algorithm:
\begin{enumerate}
  \item The respective foil counterparts for types of terms, scoped terms, and patterns,
  as well as their constructors, gain a \haskell{Foil} prefix:
  \item The types of terms and scoped terms gain an extra phantom type parameter, tracking the scope:
  \begin{itemize}
    \item \haskell{Term} $\longrightarrow$ \haskell{FoilTerm n}
    \item \haskell{ScopedTerm} $\longrightarrow$ \haskell{FoilScopedTerm n}
  \end{itemize}
  \item The types of patterns gain two extra phantom type parameters, first one corresponds to the ``outer'' scope,
  before introducing variables bound the pattern, and second one corresponds to the ``inner'' scope:
  \begin{itemize}
    \item \haskell{Pattern} $\longrightarrow$ \haskell{FoilPattern n l}
  \end{itemize}
  \item Variable types, occurring in terms and scoped terms are replaced with \haskell{Name n}
  where the scope parameter \haskell{n} is taken from the context, for example:
  \begin{itemize}
    \item \haskell{Var :: VarIdent -> Term} \\ $\longrightarrow$ \haskell{FoilVar :: Name n -> Term n}
  \end{itemize}
  \item Scoped terms are replaced with their foil counterpart with a distinct scope parameter,
  that matches the ``inner'' scope of the pattern, introduced in the same syntax constructor:
  \begin{itemize}
    \item \haskell{Lam :: Pattern -> ScopedTerm -> Term} \\
    turns into
\begin{minted}{haskell}
FoilLam :: FoilPattern n l
        -> FoilTerm l
        -> FoilTerm n
\end{minted}
    \item
\begin{minted}{haskell}
Pi :: Pattern
   -> Term
   -> ScopedTerm
   -> Term
\end{minted}
turns into
\begin{minted}{haskell}
FoilPi :: FoilPattern n l
       -> FoilTerm n
       -> FoilTerm l
       -> FoilTerm n
\end{minted}
  \end{itemize}
  \item Variables and pattern types, occurring in patterns are replaced with \haskell{Binder n l} and the generated type \haskell{FoilPattern n l} respectively
  where \haskell{n} is taken as the output scope of the previous variable or a pattern,
  and \haskell{l} is a fresh type variable. The overall type of a foil pattern constructor
  takes the first and last scope parameters as its own. Some generated data is presented in~\cref{figure:foil-data}.
  Note that \haskell{FoilPatternWildcard} has type \\ \haskell{FoilPattern n n},
  which means that scope is unchanged by this pattern (indeed, a wildcard pattern does not introduce any variables in scope!).
\end{enumerate}

The conversion functions are straightforward:
\begin{enumerate}
  \item To convert from na\"ive to scope-safe representation,
  we ask for a renaming function (to convert free variables into \haskell{Name n})
  as well as a scope. For closed terms, renaming function should always throw an error
  (since we do not expect any free variables), and an empty scope is used.
  The type signatures of the generated functions are shown in~\cref{figure:tofoil-functions}.

  \item To convert from scope-safe back to na\"ive representation,
  we simply forget the type-level scope information and convert \haskell{RawName}
  (which is the underlying representation for names used by the foil) back to original names.
  The type signatures of the generated functions are shown in~\cref{figure:fromfoil-functions}.
\end{enumerate}

\begin{figure*}
\begin{minted}{haskell}
toFoilTerm       :: Distinct n => (VarIdent -> Name n) -> Scope n -> Term       -> FoilTerm n
toFoilPattern    :: Distinct n => (VarIdent -> Name n) -> Scope n -> Pattern    -> FoilPattern n l
toFoilScopedTerm :: Distinct n => (VarIdent -> Name n) -> Scope n -> ScopedTerm -> FoilScopedTerm n
\end{minted}
\caption{Signatures of functions generated via \haskell{mkToFoil}.}
\label{figure:tofoil-functions}
\end{figure*}

\begin{figure*}
\begin{minted}{haskell}
fromFoilTerm       :: (RawName -> VarIdent) -> FoilTerm n       -> Term
fromFoilPattern    :: (RawName -> VarIdent) -> FoilPattern n l  -> Pattern
fromFoilScopedTerm :: (RawName -> VarIdent) -> FoilScopedTerm n -> ScopedTerm
\end{minted}
\caption{Signatures of functions generated via \haskell{mkFromFoil}.}
\label{figure:fromfoil-functions}
\end{figure*}

% \begin{figure*}
% \begin{minted}{haskell}
% import Language.LambdaPi.Foil.TH
% ...
% mkFoilData ''Term ''VarIdent ''ScopedTerm ''Pattern
% mkToFoil ''Term ''VarIdent ''ScopedTerm ''Pattern
% mkFromFoil ''Term ''VarIdent ''ScopedTerm ''Pattern
% mkInstancesFoil ''Term ''VarIdent ''ScopedTerm ''Pattern
% \end{minted}
% \caption{Foil interface generation.}
% \label{figure:haskell-foil-generation}
% \end{figure*}

\subsection{From EBNF to the Foil}

\begin{figure*}
\begin{minted}{ebnf}
comment "--" ;
comment "{-" "-}" ;

layout toplevel ;

token VarIdent letter (letter | digit | '_' | '\'')* ;

AProgram. Program ::= [Command] ;

CommandCheck.   Command ::= "check"   Term ":" Term ;
CommandCompute. Command ::= "compute" Term ":" Term ;
terminator Command ";" ;

Pair. Term  ::= "(" Term "," Term ")" ;
Lam.  Term ::= "lam" Pattern "." ScopedTerm ;
Pi.   Term ::= "fun" "(" Pattern ":" Term ")" "->" ScopedTerm ;
App.  Term1 ::= Term1 Term2 ;
Var.  Term2 ::= VarIdent ;
coercions Term 2 ;

AScopedTerm. ScopedTerm ::= Term ;

PatternWildcard.  Pattern ::= "_" ;
PatternVar.       Pattern ::= VarIdent ;
PatternPair.      Pattern ::= "(" Pattern "," Pattern ")" ;
\end{minted}
\caption{$\lambda\Pi$ labelled BNF grammar for the BNF Converter, with pairs, patterns, and commands around terms.}
\label{figure:lambda-pi-scoped-bnfc}
\end{figure*}

\begin{figure}
\begin{minted}{haskell}
data Term
  = Lam Pattern ScopedTerm
  | Pi Pattern Term ScopedTerm
  | App Term Term
  | Var VarIdent
  | Pair Term Term

data ScopedTerm = AScopedTerm Term

data Pattern
  = PatternWildcard
  | PatternVar VarIdent
  | PatternPair VarIdent VarIdent
\end{minted}
\caption{Abstract syntax for $\lambda\Pi$ with pairs and patterns, generated by the BNF Converter from its grammar.}
\label{figure:lambda-pi-signature}
\end{figure}

The BNF Converter~\cite{BNFConverter2004} is a tool that is capable of generating
parser, pretty-printer, abstract syntax definitions, and more, from a labelled BNF grammar
of the object language. The tool supports code generation for multiple languages,
including Haskell. For the Haskell backend, it relies on the Happy parser generator\footnote{\url{https://haskell-happy.readthedocs.io/}}.
The BNF Converter is quite useful for prototype development of small languages.
Of course, the generated abstract syntax is oblivious of the binders in the object language
and constitutes what we consider a na\"ive representation in this paper.

% BNFC-meta~\cite{BNFCMeta} - is a tool capable of generating a compiler interface based on labelled BNF grammars.
% This tool is based on the previously discussed BNF Converter, but oriented on the Haskell language.
% The main advantages of this tool are the use of Template Haskell for generation, quasi-quoter,
% static grammar checks and highly specialised grammar DSL. Despite this, the tool also generates
% na¨ıve representations. We decided to use the BNF Converter tool, as it uses a more user-friendly
% insterfes for new users. However, we do not reject BNFC-meta completely and will consider a
% user-friendly implementation with the tool in the future.

Luckily, the minimal assumptions\footnote{Namely, clear separation of
syntactic categories for terms, scopes, patterns, and variable identifiers into separate types.}
we impose on a na\"ive representation to enable automatic conversion to the foil representation can be done entirely on the level of the labelled BNF.
Indeed, the grammar in~\cref{figure:lambda-pi-scoped-bnfc} introduces a simple
language, where a part of it defines terms of $\lambda\Pi$ language with pairs and patterns.

Running the BNF Converter produces, among other things, the types for the abstract syntax
(that it also uses for parsing and pretty-printing), which are shown in~\cref{figure:lambda-pi-signature}.
It is clear that these definitions comply with our expectations, so we can
immediately use Template Haskell function described above to generate the foil representation.

Thus, with very little effort, a combination of BNF Converter and our Template Haskell functions,
allows to get a scope-safe representation based on the foil, simply from a labelled BNF grammar of the language.
Moreover, converting back from the foil and using the pretty-printer allows nicely printing such terms.

\section{Benchmarks}
\label{section:benchmarks}

% \subsection{Preparing the environment}

To see if the foil or free foil induce a runtime cost,
we run Weirich's benchmark suite~\cite{lambda-n-ways}, extending it with the following implementations
of untyped lambda calculus:
\begin{enumerate}
  \item \haskell{Foil.Lazy} — lazy implementation via the foil~\cite{MaclaurinRadulPaszke2023};
  \item \haskell{Foil.Strict} — strict implementation of the above;
  \item \haskell{FreeScoped.Lazy} — lazy implementation via free scoped monads~\cite{Kudasov2024freescoped} and Bird and Paterson's de Bruijn notation as nested data types~\cite{BirdPaterson1999};
  \item \haskell{FreeScoped.Strict} — strict implementation of the above;
  \item \haskell{FreeFoil.Lazy} — lazy implementation via the free foil (this paper);
  \item \haskell{FreeFoil.Strict} — strict implementation of the above.
  \item \haskell{NBE.Foil.Strict} — strict implementation using ``delayed substitution'', via the foil;
\end{enumerate}

Additionally, we compare against some of the fastest implementations already existing in the suite:
\begin{enumerate}
  \item \haskell{Named.Unique} — lazy implementation that ensures that all bound variables remain distinct;
  \item \haskell{Named.Simple} — strict stateless implementation of Lennart.Simple, uses sets instead of lists for free variables and map instead of single substitution for renaming;
  \item \haskell{NBE.Kovacs} — lazy implementation of call-by-need normal order evaluation;
  \item \haskell{DeBrujin.Lenart} - strict implementation of de Bruijn indices;
\end{enumerate}

Benchmark groups:
\begin{enumerate}
  \item \texttt{nf} — normalization of an extremely large lambda term;
  \item \texttt{random15} and \texttt{random20} — normalization of 100 randomly-generated lambda terms with a large number of substitutions;
\end{enumerate}

The benchmarks were conducted using a MacBook Pro 2019 with a 2.4 GHz 8-core Intel Core i9 processor and 32 GB of 2667 MHz DDR4 memory.

Our extension of the Weirich's benchmark is available at \href{https://github.com/KarinaTyulebaeva/lambda-n-ways}{github.com/KarinaTyulebaeva/lambda-n-ways}.
% This repository focuses on capture-avoiding substitution and alpha-equivalence for the untyped lambda calculus
% and already includes measurements of well-known approaches such as DeBrujin indexes,
% Locally-Nameless implementations, and Named representations. We incorporated implementations for Free Scoped Monads with DeBrujin indexes proposed by Kudashov,
% the Foil developed by Maclaurin, Radul, and Paszke, and our novel implementation of Foil with Free Scoped Monads.
% We ran nf normalization benchmarks, which include normalization of large lambda calculus terms, as well as of groups of 100 randomly generated terms.

% \subsection{Analysis of the results obtained}

\begin{figure*}
  \begin{center}
  \begin{tabular}{| l || c | c | c |}
  \hline
    & \texttt{nf}, ms & \texttt{random15}, ms & \texttt{random20}, ms \\ [0.5ex]
    \hline\hline
    \texttt{Named.Unique} & $4569.0 \pm 27.4$ & $1129.0 \pm 27.5$ & $4792.0 \pm 25.1$ \\
    \hline
    \texttt{Named.Simple} & $475.1 \pm 5.3$ & $209.8 \pm 3.6$ & $697.1 \pm 11.4$ \\
    \hline
    \texttt{DeBrujin.Lenart} & \textbf{350.0 $\pm$ 25.0} & $393.0 \pm 1.0$ & $2500.0 \pm 1.0$ \\
    \hline
    \texttt{FreeScoped.Lazy} & $4217.0 \pm 52.0$ & $6.339 \pm 0.002$ & $4.35 \pm 0.55$ \\
    \hline
    \texttt{FreeScoped.Strict} & $3297.0 \pm 160.0$ & $8.574 \pm 0.927$ & $4.913 \pm 0.522$ \\
    \hline
    \texttt{FreeFoil.Lazy} & $2866.0 \pm 71.0$  & $5.324 \pm 0.003$ & $4.495 \pm 0.003$ \\ [1ex]
    \hline
    \texttt{FreeFoil.Strict} & $536.0 \pm 8.0$ & $1057.0 \pm 23.3$ & $4225.0 \pm 27.7$ \\ [1ex]
    \hline
    \texttt{Foil.Lazy} & $2206.0 \pm 107.0$ & \textbf{2.904 $\pm$ 0.113} & \textbf{2.586 $\pm$ 0.074} \\
    \hline
    \texttt{Foil.Strict} & $360.0 \pm 14.0$ & $53.43 \pm 1.97$ & $81.96 \pm 2.47$ \\
    \hline
    \hline
    \texttt{NBE.Foil.Strict} & $1.97 \pm 0.07$ & $0.195 \pm 0.005$ & $0.304 \pm 0.083$ \\
    \hline
    \texttt{NBE.Kovacs} & $0.640 \pm 0.029$ & $0.110 \pm 0.004$ & $0.109 \pm 0.003$ \\
    \hline
  %  \texttt{NBE.KovacsScoped} & $0.00070 \pm 0.00003$ & 78 & 5415 \\
  %  \hline
  \end{tabular}
  \end{center}
  \caption{Benchmark results.
  \texttt{Named.Unique} is a lazy stateful implementation by Weirich of the Barendregt convention~\cite{Barendregt1985} with \haskell{Int} identifiers.
  \texttt{Named.Simple} is a strict stateless implementation by Weirich of the Barendregt convention, very similar to the rapier~\cite{PeytonJonesMarlow2002}.
  \texttt{Lennart.DeBruijn} is a strict implementation via de Bruijn indices by Lennart Augustsson.
  \texttt{NBE.Kovacs} is an implementation by Kovacs with normalization by evaluation.}
  \label{fig:benchmark_results}
\end{figure*}

\begin{figure*}
\begin{minted}{haskell}
data ScopedClosure sig n where
  ScopedClosure :: Binder n l -> Closure sig l -> ScopedClosure sig n

data Closure sig n where
  VarC :: Name n -> Closure sig n
  Closure :: Subst (Closure sig) n o   -- Environment (values of captured variables).
          -> sig (ScopedClosure sig n) (Closure sig n)
          -> Closure sig o
\end{minted}
\caption{A sketch of a freely generated type of closures over a given signature.}
\label{figure:free-closure}
\end{figure*}

The benchmark results in~\cref{fig:benchmark_results} show that free foil outperforms free scoped monads,
while the foil outperforms the free foil. In the process of benchmarking, we have stumbled upon several observations:
\begin{enumerate}
  \item Numerous implementations in the Weirich's repository offer two versions: lazy and strict.
  According to the benchmarks stricter implementations usually outperform lazy ones (e.g. on \texttt{nf} benchmark).
  However, some benchmarks still prefer laziness (e.g. on \texttt{random*} benchmarks).
  We have found that the free foil representation limits the ability of
  controlling strictness annotations.
  This is due to the use of \haskell{fmap} when going into subterms
  and the lack of any proper way of making \haskell{fmap} strict.

  \item The strict version of the Foil converges with the \texttt{Named.Simple}
  implementation, which closely resembles the rapier~\cite{PeytonJonesMarlow2002}.
  We conclude that foil performs on par with the rapier, despite the extra layers
  of abstraction and the use of rank-2 polymorphic functions.

  \item The most performant implementations of the untyped lambda calculus rely on
  an instance of the normalization by evaluation\footnote{See \url{https://en.wikipedia.org/wiki/Normalisation_by_evaluation}.} technique, where $\lambda$-abstractions
  get converted to closures (a technique also known as ``delayed substitutions'').
  We have implemented a version of this technique for the foil, and got similarly fast results.
\end{enumerate}

Interestingly, the free foil admits a \emph{general} definition of such closures
as can be seen in~\cref{figure:free-closure}.
Although not entirely clear, we think that such definition (perhaps, under additional assumptions)
might lead to a generic normalization by evaluation algorithm for second-order abstract syntax.
However, we leave this for future work.

\section{Conclusion}
\label{section:conclusion}

We have presented two approaches to the generation of
scope-safe abstract syntax based on the foil~\cite{MaclaurinRadulPaszke2023},
making it more accessible to functional programmers.
The two approaches come with trade-offs:
\begin{enumerate}
  \item The approach based on the free scoped monads~\cite{Kudasov2024freescoped}
  mixes well with extensions in the style of data types \`a la carte~\cite{Swierstra2008}
  and, in principle, allows implementing efficient generic language-agnostic algorithms,
  based on second-order abstract syntax~\cite{FioreHur2010}, such as higher-order unification~\cite{Kudasov2023,Kudasov2024freescoped}.
  However, the extra layer of abstraction involved with it is not entirely free and comes with a
  moderate performance penalty as we show in our benchmark. A possible improvement to the performance
  may be the use of Church encodings, similar to the ones used to improve performance of free monads~\cite{Voigtlander2008}.
  Another downside of this approach is that it is not obvious how to add general support for complex patterns,
  since we would need to properly abstract \haskell{withPattern}, which goes beyond what \haskell{CoSinkable} offers.
  Finally, the free foil provides less control over the strictness compared to explicitly recursive data types
  \item The metaprogramming approach offers a nice connection with
  the data types a functional programmer is likely to start off with as well as parser generators like BNF Converter~\cite{BNFConverter2004,BNFCMeta}.
  This approach also proves more flexible as it supports arbitrary patterns.
  On the other hand, Template Haskell is inherently more error-prone and requires
  more code to handle all possible definitions of the na\"ive syntax.
  The downside of this approach is that it does not allow for generic language-agnostic algorithms.
\end{enumerate}

\onecolumn \begin{multicols}{2} % starting twocolumn mode on a new page to ensure balanced column in references
Although presented separately, the two approaches may be united if desired,
at least in the case of simple one-variable patterns.
Indeed, it suffices to factor our the recursion and variables.
However, we leave a more general mechanism with support for arbitrary patterns for future work.

Our benchmarks show that the free foil is slightly outperformed by the foil generated via Template Haskell,
which is completely natural, as the free foil introduces an extra level of indirection for each node in the abstract syntax tree.
However, the difference in performance is not critical in our opinion.

Finally, during the benchmarks, we have noticed that ``delayed substitution''
approach significantly outperforms regular implementations. We speculate that
a generalization of this approach is possible for free scoped monads, or
second-order abstract syntax in general, perhaps under some additional constraints.

\begin{acks}
We thank the reviewers of ICCQ 2024 for their thorough and detailed comments
on a previous version of this paper. The first author is grateful to Kim-Ee Yeoh
for drawing his attention to the foil at the Workshop on the Implementation of Type Systems (WITS 2023)
and to Benedikt Ahrens for pointing out the relative monads as the proper formalism
for \haskell{AST sig}.
\end{acks}

% \printbibliography

% \onecolumn \begin{multicols}{2}
\bibliographystyle{ACM-Reference-Format}
\bibliography{references}
\end{multicols}

% \appendix









\end{document}